\documentclass[11pt]{article}

\title{Lorentz invariant phenomenology of quantum gravity: Main ideas behind the model}

\author{Yuri Bonder\footnote{This work was presented at \textit{Rencontres de Moriond and GPhyS Colloquium: Gravitational Waves and Experimental Gravity} and it was done in collaboration with A. Corichi and D. Sudarsky}}

\date{Instituto de Ciencias Nucleares\\ Universidad Nacional Aut\'onoma de M\'exico\\
A. Postal 70-543, M\'exico D.F. 04510, M\'exico\\ yuri.bonder@nucleares.unam.mx}

\begin{document}

\maketitle

\begin{small}
In the past decade the phenomenology of quantum gravity has been dominated by the search of violations of Lorentz invariance. However, there are very serious arguments that led us to assume that this invariance is a symmetry in Nature. This motivated us to construct a phenomenological model describing how a Lorentz invariant granular structure of spacetime could become manifest. The proposal is fully covariant, it involves non-trivial couplings of curvature to matter fields and leads to a well defined phenomenology.
\end{small}

\vspace{1cm}

General relativity is currently the accepted theory of spacetime and gravity and its quantum version, which is still unknown, could involve a discrete structure of spacetime at microscopic (Planckian) scales. This non-trivial microstructure is generically known as spacetime granularity and the idea of studying its consequences empirically through Lorentz invariance violations (\textit{i.e.}, by looking for a preferential reference frame) has received a great deal of attention. This is essentially because a naive granularity would take its most symmetric form in a particular reference frame. However, there are very serious experimental bounds on Lorentz invariance violations \cite{Bounds} and, moreover, the radiative corrections of a quantum field theory on a granular background that induces a preferential frame would magnify the effects of this granularity to a point where they would have been already detected \cite{Collins}.

This motivates us to assume that, if a spacetime granularity exists, it respects Lorentz invariance and we investigate if there is a phenomenological way to study its consequences. Since there is no intuitive way to imagine a discrete structure of spacetime which is Lorentz invariant, the proposal \cite{QGP} is to use an analogy: Imagine a building made of cubic bricks and having, say, a pyramidal shape. Then it is possible to detect the incompatibility between the bricks and the building symmetry, for example, by looking the mismatch at the building's surface. Given that we assume that the symmetry of spacetime's building blocks is Lorentz invariance, according to the analogy, in regions where spacetime is not Lorentz invariant, namely, where the Riemann curvature tensor ($R_{abcd}$) does not vanish, it would be possible to detect the presence of this granularity. In other words, this analogy led us to assume that a Lorentz invariant spacetime granularity could manifest through couplings of matter and $R_{abcd}$. 

For simplicity we only focus on fermionic matter fields ($\psi$) and since the Ricci tensor at $x$ is determined by the matter energy-momentum tensor at $x$, to study a coupling of Ricci and the matter fields, at a phenomenological level, looks like a self-interaction. Thus we consider the Weyl tensor ($W_{abcd}$) which, loosely speaking, is $R_{abcd}$ without Ricci. In addition, in order to produce an effect that it is observable in principle, the Lagrangian coupling term involving $W_{abcd}$ and fermionic matter fields should have mass dimensions five (we set $c=\hbar=1$) so there is no need to divide it by more than one power of a mass which is taken to be proportional to Planck mass $M_{Pl}$. Since the only coupling term of these objects with dimension five vanishes \cite{QGP}, an alternative is sought. The idea is to use $\lambda^{(s)}$ and the (dimensionless) $2$-forms $X^{(s)} _{ab}$ such that
\begin{equation}\label{eigenvalue eq}
{W_{ab}}^{cd}X^{(s)} _{cd}=\lambda^{(s)}X^{(s)} _{ab}.
\end{equation}
Observe that $s$ labels the different eigenvalues and eigen-forms of the Weyl tensor and it runs from $1$ to $6$. The first interaction Lagrangian (for one fermionic field) to be proposed \cite{QGP} is
\begin{equation}
\mathcal{L}_f= \bar{\psi} \gamma^{a}\gamma^{b} \psi \sum_{s}\frac{\xi^{(s)}}{M_{Pl}}\lambda^{(s)} X^{(s)}_{ab},
\end{equation}
where $\xi^{(s)}$ are free dimensionless parameters and $\gamma^a$ are Dirac matrices. Note that this interaction is fully covariant, however, it suffers from some ambiguities which have been cured \cite{QGP,QGP2} and that are briefly described:
\begin{itemize}
\item \textbf{Normalization}: The norm of the Weyl tensor's eigen-forms is not set by equation (\ref{eigenvalue eq}), thus, an additional condition to fix it must be given. The proposal is to use a pseudo-Riemannian metric on the space of $2$-forms that can be constructed from spacetime metric \cite{QGP}. The null eigen-forms are discarded since there is no way to normalize them, the rest are normalized to $\pm1$. 
\item \textbf{Degeneration}: The symmetries of the Weyl tensor imply that there is an unavoidable degeneration on all its eigen-forms. In fact, if we denote the spacetime volume element by $\epsilon_{abcd}$, a generic eigen-form of the Weyl tensor, $X_{ab}$, has the same eigenvalue as ${\epsilon_{ab}}^{cd}X_{cd}$. Thus, one needs a criteria to discriminate between all the linear combinations of the degenerated $2$-forms. The suggested alternative is to use the linear combinations $Y_{ab}$ satisfying $\epsilon^{abcd}Y_{ab}Y_{cd}=0$.
\item \textbf{Sign}: Equation (\ref{eigenvalue eq}) and all the conditions listed above are insensible to the substitution of any Weyl eigen-form, $X_{ab}$, by $-X_{ab}$. Essentially, we have solved this ambiguity by introducing a new coupling term which is quadratic in the eigen-forms.
\end{itemize}

Finally, let us remark that, using the formalism of the Standard Model Extension \cite{NRH} and other approximations, we have been able to obtain the non-relativistic Hamiltonian coming from this model which can be compared with experiments. Since only polarized matter is sensible to the effects predicted by the model, it is difficult to test it empirically, however, using data of Can\`{e} et al. \cite{datos} we have put bounds on some of the model's free parameters \cite{QGP2}.

\textbf{Acknowledgments} This work was supported by CONACYT 101712 project.

\end{document}